\documentclass[conference]{IEEEtran}
\IEEEoverridecommandlockouts
\usepackage{amsmath,amsfonts}
\usepackage{algorithmic}
\usepackage{array}
\usepackage[caption=false,font=normalsize,labelfont=sf,textfont=sf]{subfig}
\usepackage{textcomp}
\usepackage{stfloats}
\usepackage{url}
\usepackage{verbatim}
\usepackage{graphicx}
\usepackage[justification=centering]{caption}
\def\BibTeX{{\rm B\kern-.05em{\sc i\kern-.025em b}\kern-.08em
    T\kern-.1667em\lower.7ex\hbox{E}\kern-.125emX}}
\usepackage{balance}
\usepackage{comment}
\newenvironment{indexterms}
  {\renewcommand\abstractname{Index Terms}\begin{abstract}}
  {\end{abstract}}
   
\begin{document}

\title{}
\title{NeRF-APT: A New NeRF Framework for Wireless Channel Prediction}


\author{\IEEEauthorblockN{Jingzhou Shen, Tianya Zhao, Yanzhao Wu, Xuyu Wang}
\IEEEauthorblockA{
Knight Foundation School of Computing and Information Sciences, Florida International University, Miami, FL 33199, US\\
Emails: jshen020@fiu.edu, tzhao010@fiu.edu, yawu@fiu.edu, xuywang@fiu.edu}
}

\maketitle

\begin{abstract}
Neural radiance fields (NeRFs) have recently attracted significant attention in the realm of wireless channel prediction, primarily due to their capacity for high-fidelity reconstructions of complex wireless measurement environments. However, the ray-tracing component of NeRF-based techniques faces obstacles in realistically representing wireless scenarios, largely because the expressive power of multilayer perceptrons (MLP) remains limited. To address this challenge, in this paper, we propose NeRF-APT, an encoder-decoder architecture integrated into the NeRF-based wireless channel prediction framework. This architecture leverages the strengths of NeRF-like models in learning environmental features while capitalizing on encoder-decoder modules’ capacity to extract critical information. Furthermore, we ultilize the attention mechanism into the skip connections between encoder-decoder structures, significantly improving contextual understanding across layers. Extensive experimental evaluations on several realistic and synthetic datasets demonstrate that our approach outperforms state-of-the-art methods in wireless channel prediction.
\end{abstract}

\begin{indexterms}
Wireless Channel Prediction, Deep Learning, Neural Radiance Field (NeRF)
\end{indexterms}

\section{Introduction}
The transition from computer vision to wireless channel prediction marks a significant technological leap, evolving from spatial tracking to a comprehensive analysis of wireless data. As wireless channel prediction methods have developed, the focus has shifted from merely estimating locations to interpreting wireless signals at these locations~\cite{wang2020indoor}. This evolution has spurred the integration of advanced computer vision techniques, which are now used to reconstruct and analyze complex environments from wireless data.

The neural radiance field (NeRF) is a novel computational framework that models complex 3D environments by synthesizing photorealistic images from sparse 2D views using deep learning techniques~\cite{mildenhall2021NeRF}. It represents a groundbreaking approach in 3D vision. Inspired by the success of NeRF, NeRF2 first introduces a neural radio frequency radiance field to determine how or what signal is received at any position based on specified transmitter positions~\cite{NeRF2}. This model utilizes two multilayer perceptron~(MLP) layers to independently predict signal attenuation and reception, ultimately computing the final signal at the receiver. Similarly, WiNeRT adopts NeRF principles to efficiently and differentiably model wireless channels by predicting propagation paths and their characteristics~\cite{orekondy2023winert}. In contrast, Lu et al. argue that NeRF2 diverged from NeRF's foundational goals and propose NeWRF, a NeRF-inspired network for predicting wireless channels that incorporates distinct wireless propagation features, even in complex indoor environments with sparse measurements~\cite{lu2024newrf}.

In the 6G communication era, interests in ray-tracing technologies have intensified, notably with advancements in reconfigurable  intelligent surfaces (RIS). For example, Yang et al. introduce a novel method that synergizes NeRF models with ray-tracing to meticulously simulate dynamic electromagnetic fields in RIS-enhanced settings~\cite{yang2024rNeRFneuralradiancefields}. It effectively captures the complexities of electromagnetic signal behaviors by incorporating initial signal conditions and environmental variables. While early research primarily focused on indoor environments, recent efforts have shifted towards reconstructing large-scale urban landscapes and developing 3D city maps. For example, Zhao et al. have improved NeRF-based frameworks to accurately develop detailed 3D urban maps using GNSS data from mobile devices~\cite{crowded}. 

Although extensive research has focused on NeRF2 in various domains, there is limited research on the expressive limitations of MLP-based ray-tracing methods in wireless domain. Primarily, MLPs struggle to capture the complex and high-dimensional dynamics of wireless signal propagation because of their inherent limitations in modeling intricate dependencies and interactions within data. This inadequacy often results in a less accurate representation of environmental features critical for effective ray tracing.

To address the above challenge, we propose NeRF-APT, a novel method that leverages the synergistic integration of NeRF2 and encoder-decoder architectures to improve the generalization capabilities of ray-tracing components. The acronym "APT" stands for \textbf{A}ttention-based \textbf{P}ooling u-ne\textbf{T}. Additionally, we incorporate an attention gate to augment the contextual understanding, thereby enhancing the ability to generalize and capture the critical information of rays effectively. The main contributions of this paper are as follows.

\begin{itemize}
    \item To the best of our knowledge, this is the first to merge an encoder-decoder structure with NeRF2 in the wireless channel prediction, combining NeRF2's capabilities with U-Net to optimize ray-tracing performance.
    \item We integrate attention mechanisms and specialized pooling in U-Net, enhancing the representation of complex spatial data and improving task performance where detailed contextual understanding is critical.
    \item Our model demonstrates superior performance across both realistic and synthetic wireless datasets, surpassing established baselines and confirming the essential nature of our model enhancements.
\end{itemize}

\begin{figure}[ht]
\includegraphics[width=\columnwidth]{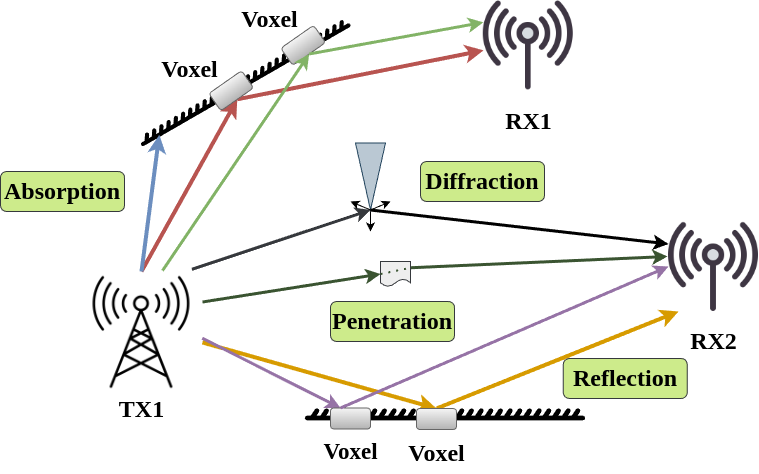}
\caption{Wireless radiance field.}
\label{sce}
\end{figure}

\section{Preliminaries}
In this section, we introduce the wireless channel prediction and the background of NeRF.

\subsection{Wireless Channel Prediction}
A wireless communication system typically comprises three key components: a transmitter, a receiver, and a communication channel. The transmitter signal represents as a complex number $X = A e^{j\theta}$, where $A$ is the amplitude, and $\theta$ is the phase. The signal generated by a transmitter~(TX), then transmits through the communication channel to an receiver~(RX). The amplitude and phase of the signal are used to modulate the transmission data. In free-space path loss model, the received signal at RX could be written as,
\begin{equation}
    Y = A e^{j\theta} \times \Delta A e^{j\Delta\theta} = A \cdot \Delta A e^{j(\theta + \Delta\theta)},
\end{equation}
where $\Delta A$ is the attenuation coefficient, and $\Delta \theta$ is the phase rotation. As shown in Fig.~\ref{sce}, the wireless radiance field is affected by reflection, absorption, diffraction, and penetration, where each voxel acts as a new radiance source retransmitting the wireless signal. Due to various wireless propagation effects, the received signal is the accumulation of multiple copied transmitted signals, which can be expressed by,
\begin{equation}
Y = A e^{j\theta} \times \sum_{l=0}^{L-1} \Delta A^l e^{j\Delta\theta_l},
\end{equation}
where $L$ is the total number of propagation paths, and $l$ is the path index. The wireless channel can be modeled as the ratio of the received signal and transmitted signal:
\begin{equation}
H = \frac{Y}{X} = \sum_{l=0}^{L-1} \Delta A^l e^{j\Delta\theta_l}.
\end{equation}
Then, the received signal strength indicator (RSSI) is defined as the received power in decibels (dB), which is formulated by $RSSI=10\log_{10}(\|H\|^2)$. Moreover, the wireless channel can be used to estimate the spatial spectrum, which represents the spatial power distribution function, with an antenna array~\cite{NeRF2}. Therefore, in this paper, our objective is to develop a new NeRF framework to accurately predict the wireless channel, received power, and spatial spectrum. 

\subsection{Neural Radiance Field}
NeRF is a novel framework for synthesizing photorealistic images from sparse sets of 2D views of a scene. NeRF employs MLPs to model the volumetric scene function $f : (x, y, z, \theta, \phi) \rightarrow (r, g, b, \sigma)$. It maps 5D coordinates (spatial locations $x, y, z$ and viewing directions $\theta$, $\phi$) to RGB color $r, g, b$ and volume density $\sigma$. To synthesize an image, NeRF effectively captures the interaction of light with the environment by integrating color and density contributions along camera rays. Each pixel in the image corresponds to a unique ray that originates from the camera's position and projects into the scene, its direction dictated by the camera's orientation and field of view. The color $\textbf{C}(\textbf{r})$ along the ray is defined by,
\begin{equation}
\textbf{C}(\mathbf{r}) = \int_{t_n}^{t_f} T(t) \sigma(\mathbf{r}(t)) \mathbf{c}(\mathbf{r}(t), \mathbf{d}) \, dt,
\end{equation}
where \( T(t) = \exp\left(-\int_{t_n}^{t} \sigma(\mathbf{r}(s)) \, ds\right) \) represents the accumulated transmittance along the ray from $t_n$ to $t_f$, where \( t_n \) and \( t_f \) represent the near and far boundaries, respectively, modeling how much light is blocked before reaching each point, \(\textbf{r}(t)\) expresses as as ray \(\textbf{r}(t) = \textbf{o} + t\textbf{d}\), where \(\textbf{o}\) is the camera origin, \(t\) is the distance along the ray and \(\textbf{d}\) is the direction of the ray. \(\mathbf{c}(\mathbf{r}(t), \mathbf{d})\) represents the color at the position of \(\mathbf{r}(t)\) and direction \(\mathbf{d}\).

However, traditional MLPs face difficulties with high-dimensional data and efficiently handling spatial or sequential information, particularly due to challenges like vanishing and exploding  gradients~\cite{glorot2010understanding}. Currently, more advanced models use complex models and better training techniques. U-Net~\cite{ronneberger2015u} performs well in complex tasks by capturing detailed spatial information~\cite{springenberg2014striving}. Thus, in this paper, we employ an U-Net structure, which selectively extracts critical information from sequence inputs, to replace MLPs in NeRF2.


\section{Framework}

\subsection{Model Structure}
\begin{figure}[htbp]
\includegraphics[width=\columnwidth]{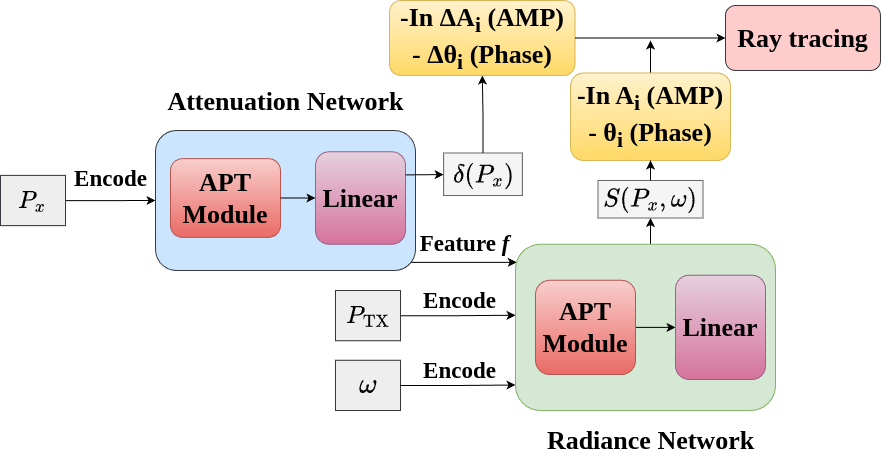}
\caption{NeRF-APT framework based on NeRF2.}
\label{figure_nu}
\end{figure}

In this section, we introduce our model, NeRF-APT, depicted in Fig.~\ref{figure_nu}. The model comprises two specialized subnetworks, which are the attenuation network and the radiance network. The attenuation network processes the encoded position of a voxel, \(P_{\text{x}}\), to output an attenuation coefficient \(\delta(P_{\text{x}})\) and a feature \(f\). This coefficient, \(\delta(P_{\text{x}})\), reflects both amplitude (AMP) degradation and phase rotation as a ray passes through \(P_{\text{x}}\), with values \(\Delta A\) and \(\Delta \theta\), respectively. It depends only on the voxel's density and the scene's structural component, making it independent of incoming wireless signals. The radiance network uses encoded information of the transmitter's position \(P_{\text{TX}}\), direction \(\omega\), and the feature \(f\) from the attenuation network to generate the retransmitted signal \(S(P_{\text{x}}, \omega)\). The embedding function is the positional encoder which combines sine and cosine functions related to the maximum frequency.

\subsubsection{The Attenuation Network}

\begin{figure}[htbp]
\includegraphics[width=\columnwidth             
                     ]{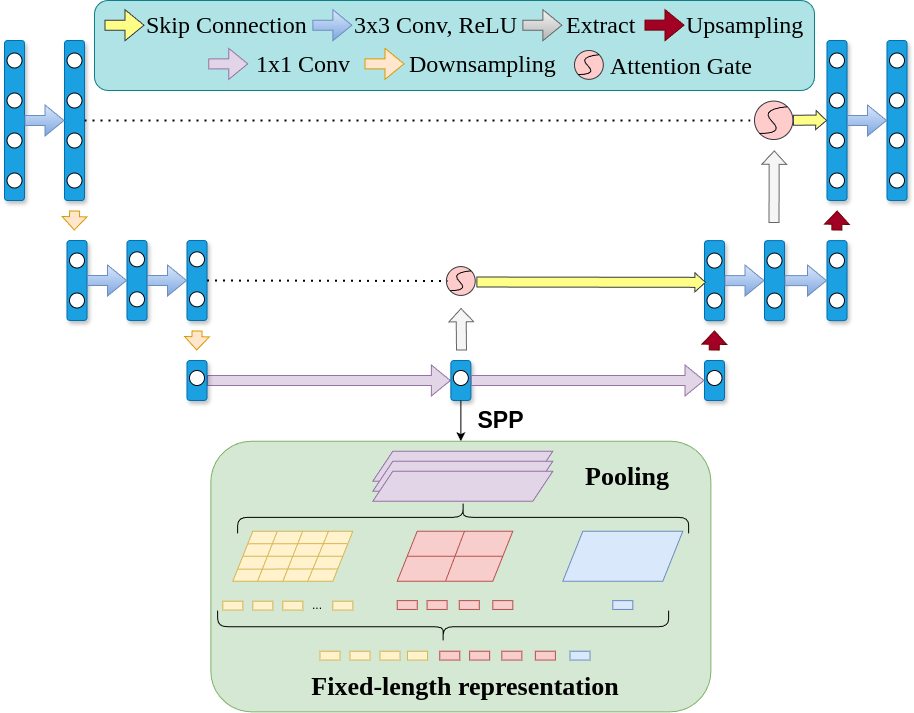}
\centering
\caption{APT structure.}
\label{u}
\end{figure}

The attenuation network is structured with two core components: a U-Net enhanced by an attention gate~\cite{vaswani2017attention} and a specialized pooling module, designated as an APT module, and a subsequent linear layer. In Fig.~\ref{u}, our APT module is strucutred with the U-Net. The U-Net architecture employs an encoder-decoder framework with skip connections, integrating high-level semantic and detailed spatial information to enable precise reconstructions and capture multi-scale features, enhancing adaptability in complex, high-resolution environments. In each intermediate layer, we perform 3$\times$3 convolution twice on the feature. Furthermore, we utilize adaptive pooling stragety to fix the dimension mismatch problem.

\begin{figure}[htbp]
\includegraphics[width=\columnwidth]{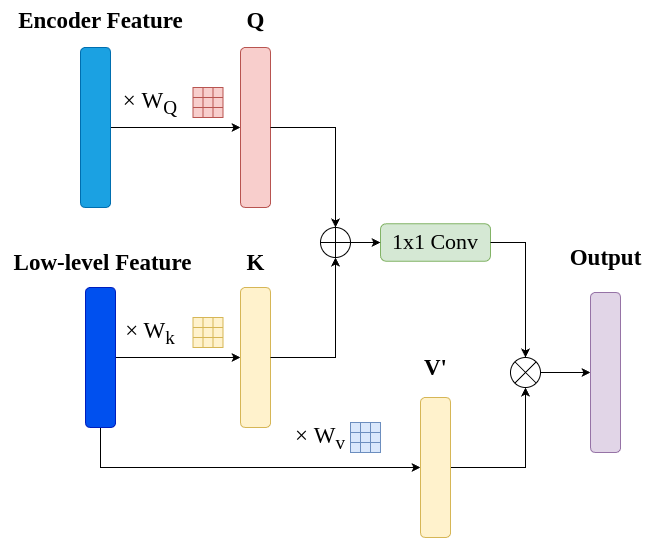}
\centering
\caption{Simplified attention gate.}
\label{att}
\end{figure}

Following the principles of attention U-Net~\cite{oktay2018attention}, we integrate an attention gate between matching encoder and decoder layers. As illustrated in Fig.~\ref{att}, the process starts with the input of low-level encoder features, which are transformed into queries ($Q$), keys ($K$), and values ($V'$) using their respective weight matrices. Initially, $Q$ and $K$ are merged, and this merged output is processed through a softmax layer followed by a 1$\times$1 convolutional layer. The output from this process is then multiplied by $V'$ to generate the final feature output. The attention mechanism can be defined by,
\begin{equation}
\text{Attention\_Gate}(Q, K, V') = \text{softmax}(\Phi\left(\frac{Q+K}{\sqrt{d_k}}\right)) V',
\end{equation}
where \( Q \) is the output of the encoder feature, weighted by matrix \( W_Q \), while both \( K \) and \( V' \) are derived from the low-level feature, using matrices \( W_K \) and \( W_V \) respectively. These low-level features originate from an intermediate layer of the previous decoder module. \( d_k \) represents the dimension of \( K \). To prevent the dot products from becoming excessively large with increasing vector dimensions, we divide by \( \sqrt{d_k} \), which helps avoid vanishing gradients and maintains model stability and effectiveness. \( \Phi \) denotes the 1$\times$1 convolution layer.

We also integrate a spatial pyramid pooling (SPP) module into the bottleneck layer of our U-Net architecture to efficiently manage high-dimensional features. The SPP module divides these features into progressively smaller subregions, applies adaptive pooling to each for capturing multi-scale information, and produces a uniform, fixed-length representation. This ensures compatibility with various input shapes and preserves essential spatial details. In practice, the input feature map undergoes parallel pooling operations at different scales, typically with kernel sizes that scale by powers of two. Each operation reduces the spatial dimensions to a set size, followed by a 1$\times$1 convolution that adjusts the channel dimensions to match the input. The resulting features are concatenated and sent to the next layer.

By integrating attention mechanisms and a pooling strategy into the U-Net architecture, we can better extract key features from sample points. This improves the model's ability to infer unknown voxel information from known data along a ray, while also enhancing its interpretability and performance.

For enhanced computational efficiency, a logarithmic transformation of the amplitude coefficient is applied to the output, which shows as follows:
\begin{equation}
    \begin{aligned}
\delta (V_i) &= -\ln(\Delta A(V_i) e^{j\Delta \theta(V_i)}) \\
             &= -\ln \Delta A(V_i) - j\Delta \theta(V_i),
\end{aligned}
\end{equation}
where $V_i$ points out the i-th voxel along the ray.

\subsubsection{The Radiance Network}

A crucial material property feature \textit{f} is extracted from the attenuation network. This feature, along with $P_{\text{TX}}$ and direction $\omega$, is fed into the radiance network. The direction $\omega$ includes both azimuthal and elevation angles. The radiance network has the same design as the attenuation network, which consists of an APT and a linear module, and further generates the re-transmitted signal. $S(P_{\text{x}}, \omega)$ in Eq.~\ref{eq}. In addition, a logarithmic transformation of the amplitude is also applied, as shown in Fig.~\ref{figure_nu}.
\begin{equation}
    S(P_{\text{x}}, \omega) = A(P_{\text{x}}) e^{j \theta(P_{\text{x}})}.
\label{eq}
\end{equation}

\subsection{Ray-Tracing}
The outputs of these subnetworks collectively inform the ray-tracing module, enabling it to compute the corrected complex signal values. The voxel along this ray could be defined as following in ray-tracing:
\begin{equation}
    P(r, \omega) = P_{\text{RX}} + r_\omega,
\end{equation}
where $r_\omega$ is the radial distance from the RX to the point on the ray from direction \( \omega \).

In our coordinate system, we define the received power at the location of the receiver as \( P_{\text{RX}} = P(0, \omega) \). We collect signals from each voxel and subsequently aggregate the re-transmitted signals from all directions. This aggregation is mathematically represented by the following equation~\cite{NeRF2}:
\begin{equation}
    R(\omega) = \int_0^D H_{P(r,\omega) \rightarrow P_{\text{RX}}} S(P(r, \omega), -\omega) \, dr,
\label{rrr}
\end{equation}
where \( H_{P(r,\omega) \rightarrow P_{\text{RX}}} \), measures the signal loss as it travels from \( P(r,\omega) \) to \( P_{\text{RX}} \), accounting for any medium or distance-related reductions. The term \( S(P(r,\omega), -\omega) \) represents the signal sent from the voxel at \( P(r,\omega) \) towards the receiver at \( P_{\text{RX}} \), with \( -\omega \) indicating transmission in the direction opposite to the ray’s travel. The variable \( D \) specifies the scene’s maximum dimension, setting the integral's upper boundary and thus the scope of the analysis.
    
In practical applications, we often rely on actual sampled measurements rather than a continuous function. Assuming uniform distributions along each optical path, the APT module effectively models the re-transmitted signals received at each voxel. As signals travel toward the receiver, they are attenuated by intervening voxels. Therefore, ray tracing in a specific direction entails aggregating re-transmitted signals from all voxels along the ray, with each voxel serving as a distinct emission source. This process can be discretized by,
\begin{equation}
\begin{aligned}
R_{\text{ray}}(\omega) &= \sum_{n=1}^N R_{\text{voxel}}(V_n) \\
&= \sum_{n=1}^N \left( \exp \left( -\sum_{m=1}^{n-1} \delta(V_m) \right) \cdot R(V_n, -\omega) \right),
\label{rwe}
\end{aligned}
\end{equation}
where $V$ denotes as voxels, $V_1$ being the closest to the receiver and $V_N$ the farthest. $\delta(V_m)$ is the attenuation coefficient at voxel $V_m$ , which is located between the RX and the current voxel $V_n$, and $N$ is the total number of voxels. \(R(V_n, -\omega)\) denotes the re-transmitted signal at the n-th voxel in the direction of $-\omega$.

\section{Experiments and Results}
\subsection{Datasets}
In our paper, we analyze both real-world and synthetic datasets. We classify the real-world datasets into three categories based on the NeRF2 framework: (1) a BLE-RSSI gateway dataset from an elderly nursing facility, (2) an RFID dataset, and (3) a MIMO-CSI dataset using the Argos channel data~\cite{NeRF2}.

The BLE-RSSI dataset comprises fifty BLE gateways, each outfitted with an NRF52832 Bluetooth SoC chip from Nordic Semiconductor. These gateways are strategically positioned throughout the facility and operate at 2.4 GHz. They record IDs and RSSI values from BLE beacons embedded in visitor badges and wristbands. To enhance signal reliability, additional gateways are deployed to achieve triple coverage in every area.

The RFID dataset is created using a USRP-based receiver and a 4$\times$4 antenna array, capturing data from a mobile RFID tag at various locations to increase the diversity of the measurement scenarios.

The MIMO-CSI dataset is collected with the ArgosV2 platform~\cite{argos}, which uses omnidirectional monopole antennas spaced at half a wavelength for 2.4 GHz, supporting up to 20 MHz bandwidth and 64 orthogonal frequency-division multiplexing (OFDM) subcarriers. Channel state information (CSI) values are obtained by transmitting 802.11 long training symbol pilots at the beginning of each frame.

For synthetic data, we use the public NewRF code~\cite{lu2024newrf} to generate different wireless channel datasets in three 3D environments in MATLAB: a bedroom, a conference room, and an office. We enhance the synthetic data by incorporating an OFDM system with transmitters randomly placed in each scene. Antenna positions are specified, and the MATLAB Antenna Toolbox is employed to generate the datasets. To prevent overfitting, we applied cross-validation, allocating 80\% of the data for training and 20\% for evaluation.

\begin{figure}
  \centering
  \begin{tabular}{@{}c@{}}
    \includegraphics[width=\columnwidth]{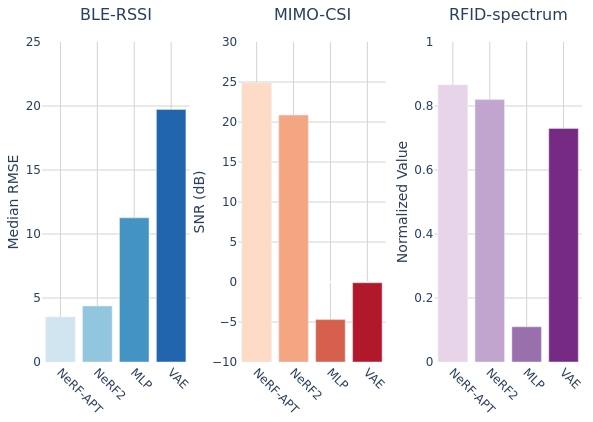} \\[\abovecaptionskip]
    \small a) Results on real-world datasets.
  \end{tabular}

  \vspace{\floatsep}

  \begin{tabular}{@{}c@{}}
    \includegraphics[width=\columnwidth]{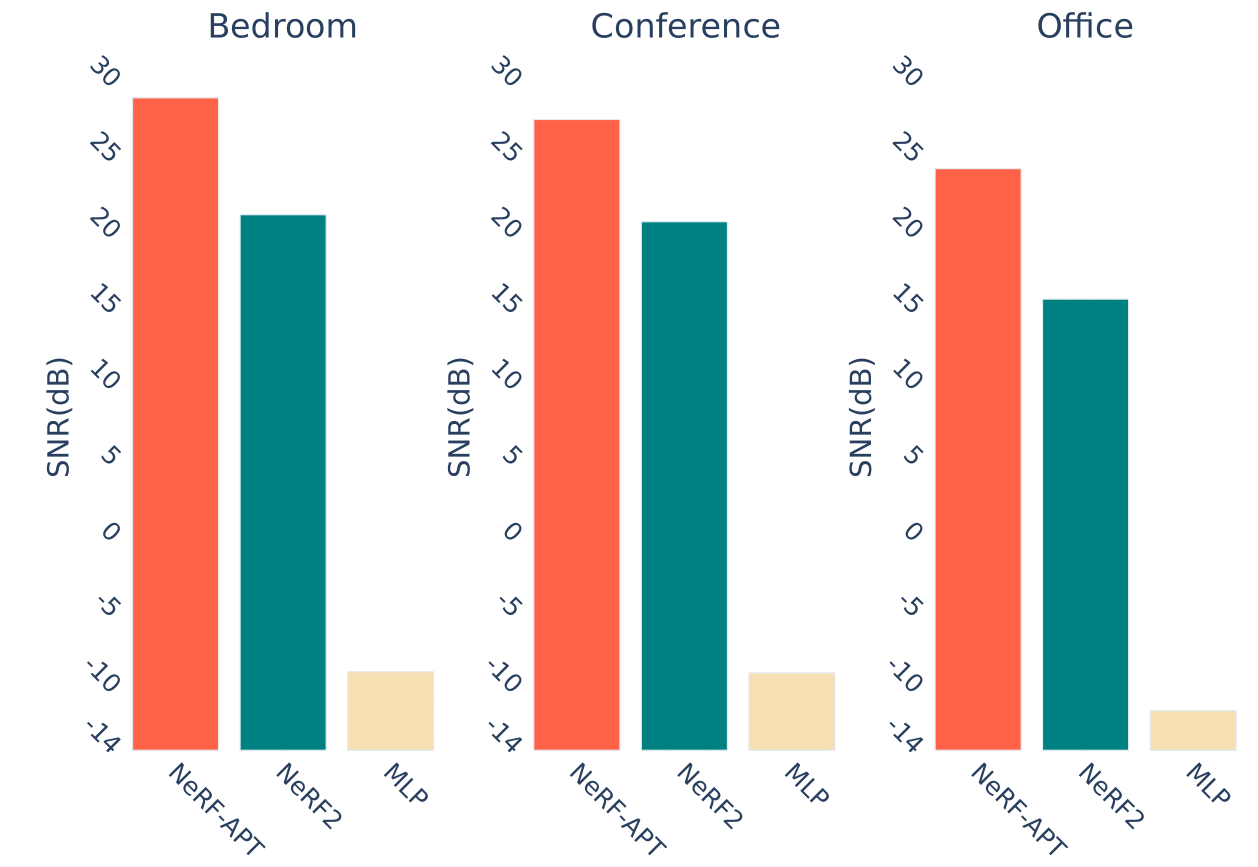} \\[\abovecaptionskip]
    \small b) Results on synthetic datasets.
  \end{tabular}

  \caption{Experiment results.}\label{fig:myfig}
\end{figure}

\subsection{Results}
We conduct experiments using our model and several baselines within the neural network intelligence framework, an open-source AutoML toolkit known for its efficient hyperparameter tuning capabilities~\cite{nni2021}. Fig.~\ref{fig:myfig} shows the results from both real-world and synthetic datasets. Here, we demonstrate the effectiveness and robustness of our model by comparing its performance with various baselines across different test environments.

Fig.~\ref{fig:myfig}(a) shows our results using real-world datasets. We compare our model with NeRF2, MLP and variational autoencoder~(VAE)~\cite{kingma2013auto}. Specifically, in the BLE-RSSI dataset, we use a median root MSE~(RMSE) as the metric, which is defined as $\text{RMSE} = \sqrt{\frac{1}{n} \sum_{i=1}^n (\hat{y}_i - y_i)^2}$, where $\hat{y}$ and $y$ represent the measured and actual RSSI, respectively. Our model demonstrates superior performance, achieving a median RMSE error of 3.52 dB, which has an improvement over NeRF2’s 4.37 dB and significantly better than the VAE's error of 11.26 dB and MLP’s error of 19.73 dB. 

For the MIMO-CSI dataset, we utilize signal-to-noise ratio~(SNR) as our evaluation criterion, which is defined by, 
\begin{equation}
SNR = 10 \log_{10} \left(\frac{P_{\text{signal}}}{P_{\text{noise}}}\right).
\end{equation}
SNR measures signal strength relative to background noise, thus providing a critical indication of communication quality and system reliability. In this evaluation, our model achieves an SNR of 24.91 dB, outperforming NeRF2’s 20.87 dB and markedly surpassing MLP’s -4.70 dB and VAE's -0.09 dB.

In the analysis of the RFID-spectrum dataset, we apply the SSIM as our metric, which define as the following:
\begin{equation}
    \text{SSIM}(x,y) = \frac{(2\mu_x \mu_y + c_1)(2\sigma_{xy} + c_2)}{(\mu_x^2 + \mu_y^2 + c_1)(\sigma_x^2 + \sigma_y^2 + c_2)},
\end{equation}
where \(\mu_x\) and \(\mu_y\) are the average of \(x\) and \(y\). \(\sigma_x^2\) and \(\sigma_y^2\) are the variance of \(x\) and \(y\). \(\sigma_{xy}\) is the covariance of \(x\) and \(y\). \(c_1 = (k_1 L)^2\), \(c_2 = (k_2 L)^2\) are two variables to stabilize the division with weak denominator. \(L\) is the dynamic range of the pixel-values, \(k_1 = 0.01\) and \(k_2 = 0.03\) by default.

\begin{figure}[htp]
    \centering
    \begin{tabular}{ccc}
        \includegraphics[scale=0.22]{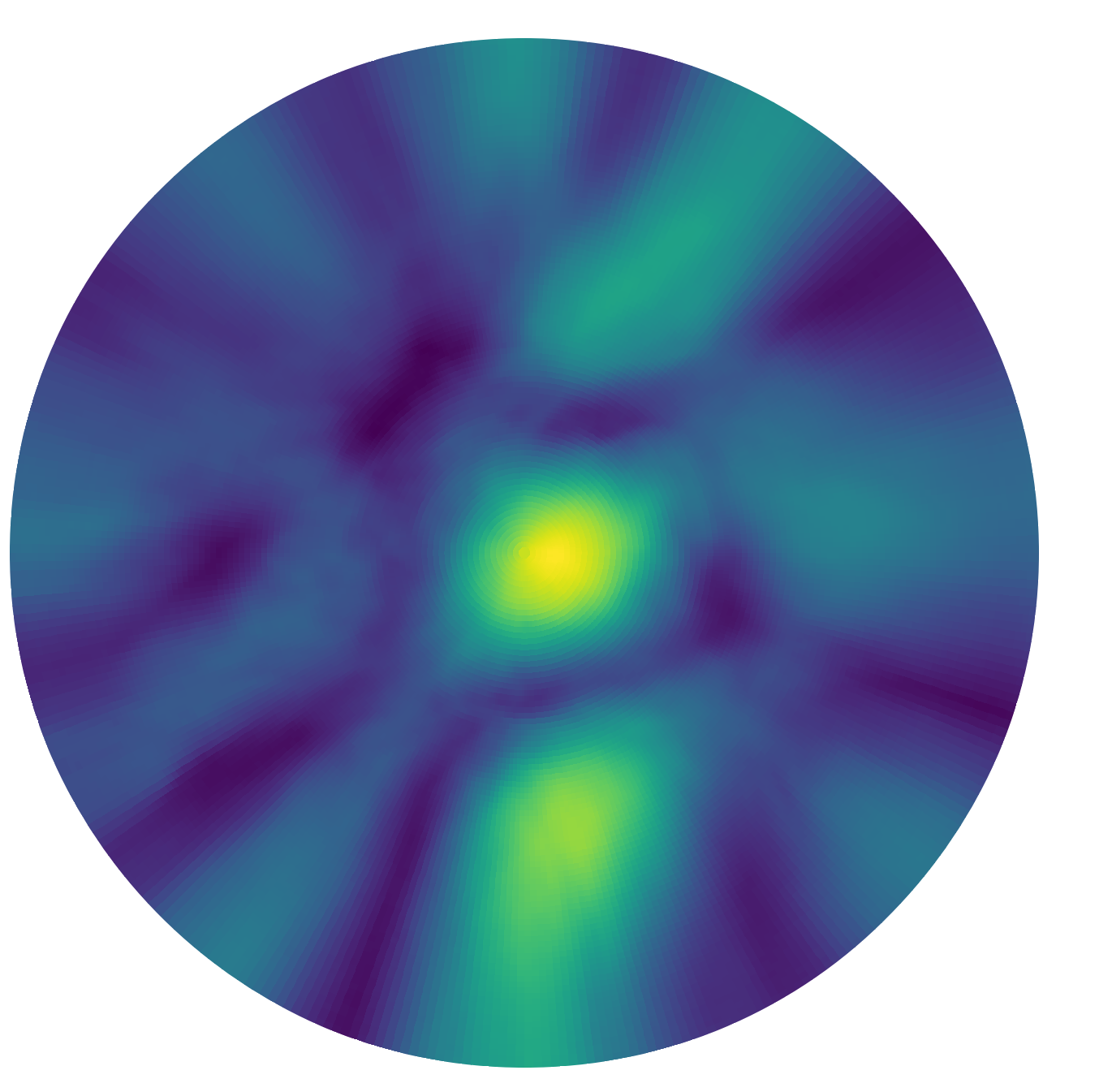} &
        \includegraphics[scale=0.22]{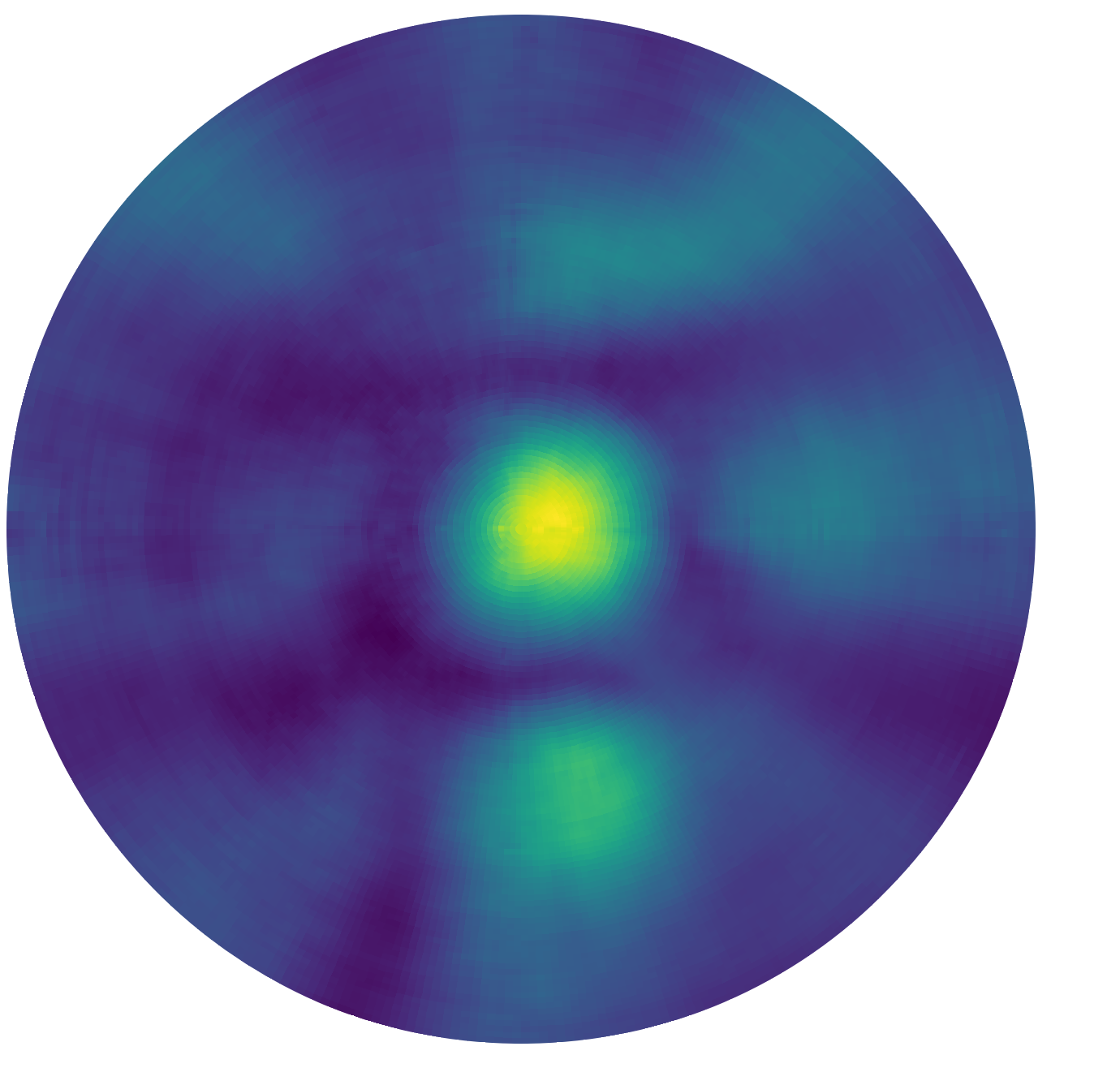} &
        \includegraphics[scale=0.22]{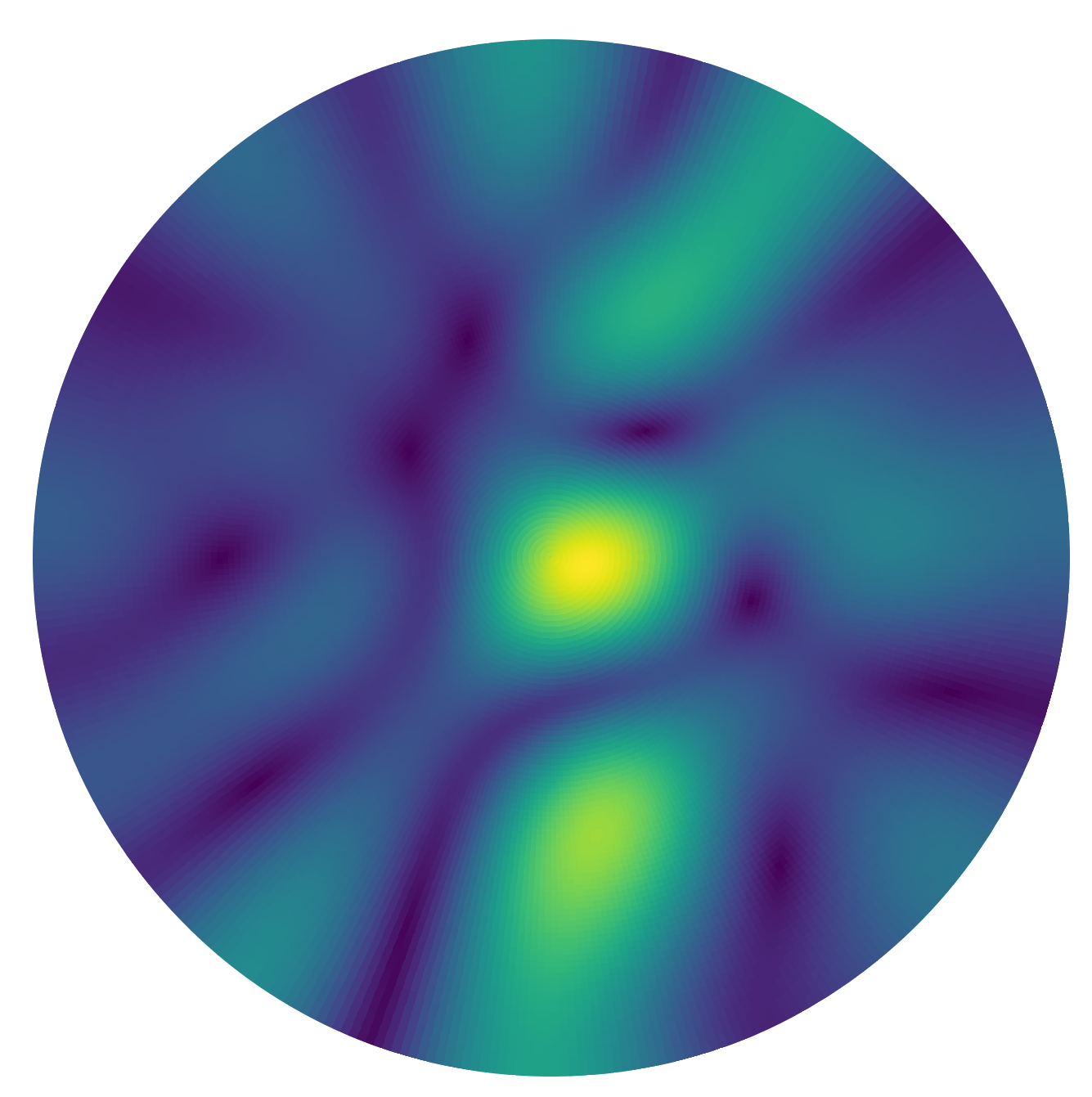} \\
        (a) NeRF-APT & (b) NeRF2 & (c) Label
    \end{tabular}
    \caption{Results on RFID-spectrum dataset.}
    \label{fig:tabularimages}
\end{figure}

SSIM assesses the perceptual quality of spectrum images by evaluating their structural similarity, thus offering a robust measure of visual fidelity. Our methodology attains an SSIM of 0.866, surpassing NeRF2’s 0.82 and greatly excelling MLP’s 0.11 and VAE's 0.73. Additionally, Fig.~\ref{fig:tabularimages} illustrates a range of data and comparative results from our model and NeRF2, clearly evidencing our model’s enhanced ability to accurately map the spatial distribution and intensities of signals.

Fig.~\ref{fig:myfig}(b) shows the better performance of our model on synthetic datasets, where it consistently surpasses both baseline models across a range of scenarios. We compare our model with NeRF2, NewRF and MLP. Specifically, in a bedroom environment, our model achieves a SNR of 29.06 dB, significantly outperforming NeRF2, which attains an SNR of 21.34 dB, and greatly exceeding the MLP's SNR of -8.81 dB and NewRF's SNR of -2.19 dB. In a conference room setting, our method records an SNR of 27.13 dB, compared to NeRF2's 20.88 dB, MLP's -8.89 dB, and NewRF's -0.9548 dB. Similarly, within an office environment, our model obtains an SNR of 24.38 dB, notably better than NeRF2’s 15.77 dB and a substantial improvement over MLP's -11.38 dB.

The original NewRF paper raises concerns that NeRF2 does not adhere strictly to the traditional NeRF method. Despite numerous iterations, NewRF seems tailored for environments limited to a single channel. When operating with data from an OFDM system, NewRF's performance deteriorates markedly. One reason is that the sampling and optimization strategies utilized in NewRF are not well-adapted for OFDM systems, resulting in the extracted coarse weights converging towards zero. Furthermore, NewRF modifies the positions of TX and RX to align with the traditional NeRF setup. We have tested both configurations in our model and found that the performance remains comparable in OFDM systems. 

\renewcommand{\arraystretch}{1.33}
\captionsetup[table]{skip=10pt} 
\begin{table}[h]
    \centering
    \begin{tabular}{|p{2.8 cm}|c|c|c|}
    \hline
    & \textbf{BLE-RSSI} & \textbf{MIMO-CSI} \\
    & \textbf{(Median RMSE / dB)} & \textbf{(SNR / dB)}\\
    \hline
    NeRF-U-Net (1 Layer) & 3.62 & 24.37 \\
    NeRF-U-Net (2 Layers) & 3.52  & 24.35 \\
    NeRF-U-Net (3 Layers) & 3.63 & 24.59\\
    NeRF-APT (2 Layers) & \textbf{3.51} & 24.62 \\
    NeRF-APT (3 Layers) & 3.56 & \textbf{24.91} \\
    \hline
    \end{tabular}
    \caption{Performance of different structures of NeRF-APT in real-world datasets.}
    \label{tab:performance}
\end{table}

We demonstrate how our model's performance varies across different settings in various tasks. The two tables highlight the importance of the APT module and different U-Net structures. As shown in Table~\ref{tab:performance}, using two layers in a U-Net architecture and adding our APT module significantly affects performance. The BLE-RSSI dataset shows optimal performance with just changes to the U-Net structure. With two layers, the NeRF-APT model achieves a median RMSE of 3.51 dB, surpassing all other baselines. However, with the MIMO-CSI dataset, the best SNR of 24.91 dB is achieved using a three-layer NeRF-APT configuration.

For synthetic datasets, as depicted in Table.~\ref{tab:performance_s}, the NeRF-APT model with three layers exhibits outstanding performance in both bedroom and conference scenarios, achieving scores of 29.06 dB and 27.63 dB, while two-layer NeRF-APT model peaks at 24.38 dB in the office scene.

\renewcommand{\arraystretch}{1.4}
\captionsetup[table]{skip=10pt} 
\begin{table}[h]
    \centering
    \begin{tabular}{|p{2.8cm}|c|c|c|}
    \hline
    & \textbf{Bedroom} & \textbf{Conference} & \textbf{Office} \\
    & \textbf{(SNR / dB)} & \textbf{(SNR / dB)} & \textbf{(SNR / dB)} \\
    \hline
    NeRF-U-Net (1 Layer) & 23.81 & 25.95 & 19.41\\
    NeRF-U-Net (2 Layers) & 14.97 & 26.63 & 20.79\\
    NeRF-U-Net (3 Layers) & 21.66 & 24.62 & 23.23\\
    NeRF-APT (2 Layers) & 25.47 & 26.75 & \textbf{24.38}\\
    NeRF-APT (3 Layers) & \textbf{29.06} & \textbf{27.63} & 24.13\\
    \hline
    \end{tabular}
    \caption{Performance of different structures of NeRF-APT in synthetic datasets.}
    \label{tab:performance_s}
\end{table}

The table reveals the peak performance achieved by various configurations of the NeRF-APT structure across different tasks. The optimal structure for NeRF-APT varies among tasks, primarily due to two factors. First, the required dimensions of feature maps vary for each task, requiring adjustments in the layer depth to optimize performance. Second, some tasks require adaptive pooling to avoid size mismatches during upsampling, which could otherwise degrade performance. Tailoring the NeRF-APT structure to these specific requirements is crucial for improving its effectiveness across various applications.

\section{Conclusion}
In this paper, we propose the NeRF-APT model, which combines an encoder-decoder structure with NeRF2 for wireless channel prediction. Our model features an enhanced bottleneck layer that integrates an SPP module, significantly improving the model's ability to capture multi-scale contextual information. Furthermore, we incorporate an attention gate within this layer, enhancing the model's capacity to identify long-range dependencies and complex interactions among sampled points. This improvement strengthens representation learning and the quality of the generated rays. Our extensive evaluations on both real-world and synthetic datasets yield promising results, validating the effectiveness of our approach.
\section*{Acknowledgments}
This work is supported in part by the NSF (CNS-2415209, CNS-2321763, CNS-2317190, IIS-2306791, and CNS-2319343).

\bibliographystyle{IEEEtran}
\bibliography{ref}

\begin{thebibliography}{10}
\providecommand{\url}[1]{#1}
\csname url@samestyle\endcsname
\providecommand{\newblock}{\relax}
\providecommand{\bibinfo}[2]{#2}
\providecommand{\BIBentrySTDinterwordspacing}{\spaceskip=0pt\relax}
\providecommand{\BIBentryALTinterwordstretchfactor}{4}
\providecommand{\BIBentryALTinterwordspacing}{\spaceskip=\fontdimen2\font plus
\BIBentryALTinterwordstretchfactor\fontdimen3\font minus \fontdimen4\font\relax}
\providecommand{\BIBforeignlanguage}[2]{{%
\expandafter\ifx\csname l@#1\endcsname\relax
\typeout{** WARNING: IEEEtran.bst: No hyphenation pattern has been}%
\typeout{** loaded for the language `#1'. Using the pattern for}%
\typeout{** the default language instead.}%
\else
\language=\csname l@#1\endcsname
\fi
#2}}
\providecommand{\BIBdecl}{\relax}
\BIBdecl

\bibitem{wang2020indoor}
X.~Wang, X.~Wang, S.~Mao, J.~Zhang, S.~C. Periaswamy, and J.~Patton, ``Indoor radio map construction and localization with deep {G}aussian processes,'' \emph{IEEE Internet of Things Journal}, vol.~7, no.~11, pp. 11\,238--11\,249, 2020.

\bibitem{mildenhall2021NeRF}
B.~Mildenhall, P.~P. Srinivasan, M.~Tancik, J.~T. Barron, R.~Ramamoorthi, and R.~Ng, ``Nerf: Representing scenes as neural radiance fields for view synthesis,'' \emph{Communications of the ACM}, vol.~65, no.~1, pp. 99--106, 2021.

\bibitem{NeRF2}
X.~Zhao, Z.~An, Q.~Pan, and L.~Yang, ``Nerf2: Neural radio-frequency radiance fields,'' in \emph{Proceedings of the 29th Annual International Conference on Mobile Computing and Networking}, 2023, pp. 1--15.

\bibitem{orekondy2023winert}
T.~Orekondy, P.~Kumar, S.~Kadambi, H.~Ye, J.~Soriaga, and A.~Behboodi, ``Wi{N}e{RT}: Towards neural ray tracing for wireless channel modelling and differentiable simulations,'' in \emph{The Eleventh International Conference on Learning Representations}, 2023.

\bibitem{lu2024newrf}
\BIBentryALTinterwordspacing
H.~Lu, C.~Vattheuer, B.~Mirzasoleiman, and O.~Abari, ``Ne{WRF}: A deep learning framework for wireless radiation field reconstruction and channel prediction,'' in \emph{Forty-first International Conference on Machine Learning}, 2024. [Online]. Available: \url{https://openreview.net/forum?id=181hXof7ho}
\BIBentrySTDinterwordspacing

\bibitem{yang2024rNeRFneuralradiancefields}
\BIBentryALTinterwordspacing
H.~Yang, Z.~Jin, C.~Wu, R.~Xiong, R.~C. Qiu, and Z.~Ling, ``{R-NeRF}: Neural radiance fields for modeling ris-enabled wireless environments,'' 2024. [Online]. Available: \url{https://arxiv.org/abs/2405.11541}
\BIBentrySTDinterwordspacing

\bibitem{crowded}
X.~Zhao, S.~Wang, Z.~An, and L.~Yang, ``Crowdsourced geospatial intelligence: Constructing {3D} urban maps with satellitic radiance fields,'' \emph{Proceedings of the ACM on Interactive, Mobile, Wearable and Ubiquitous Technologies}, vol.~8, no.~3, pp. 1--24, 2024.

\bibitem{glorot2010understanding}
X.~Glorot and Y.~Bengio, ``Understanding the difficulty of training deep feedforward neural networks,'' in \emph{Proceedings of the thirteenth international conference on artificial intelligence and statistics}.\hskip 1em plus 0.5em minus 0.4em\relax JMLR Workshop and Conference Proceedings, 2010, pp. 249--256.

\bibitem{ronneberger2015u}
O.~Ronneberger, P.~Fischer, and T.~Brox, ``U-net: Convolutional networks for biomedical image segmentation,'' in \emph{Medical image computing and computer-assisted intervention--MICCAI 2015: 18th international conference, Munich, Germany, October 5-9, 2015, proceedings, part III 18}.\hskip 1em plus 0.5em minus 0.4em\relax Springer, 2015, pp. 234--241.

\bibitem{springenberg2014striving}
J.~T. Springenberg, A.~Dosovitskiy, T.~Brox, and M.~Riedmiller, ``Striving for simplicity: The all convolutional net,'' \emph{arXiv preprint arXiv:1412.6806}, 2014.

\bibitem{vaswani2017attention}
A.~Vaswani, ``Attention is all you need,'' \emph{Advances in Neural Information Processing Systems}, 2017.

\bibitem{oktay2018attention}
O.~Oktay, J.~Schlemper, L.~L. Folgoc, M.~Lee, M.~Heinrich, K.~Misawa, K.~Mori, S.~McDonagh, N.~Y. Hammerla, B.~Kainz \emph{et~al.}, ``Attention u-net: Learning where to look for the pancreas,'' \emph{arXiv preprint arXiv:1804.03999}, 2018.

\bibitem{argos}
\BIBentryALTinterwordspacing
C.~Shepard, H.~Yu, N.~Anand, E.~Li, T.~Marzetta, R.~Yang, and L.~Zhong, ``Argos: practical many-antenna base stations,'' in \emph{Proceedings of the 18th Annual International Conference on Mobile Computing and Networking}, ser. Mobicom '12.\hskip 1em plus 0.5em minus 0.4em\relax New York, NY, USA: Association for Computing Machinery, 2012, p. 53–64. [Online]. Available: \url{https://doi.org/10.1145/2348543.2348553}
\BIBentrySTDinterwordspacing

\bibitem{nni2021}
\BIBentryALTinterwordspacing
{Microsoft}, ``{Neural Network Intelligence},'' 1 2021. [Online]. Available: \url{https://github.com/microsoft/nni}
\BIBentrySTDinterwordspacing

\bibitem{kingma2013auto}
D.~P. Kingma, ``Auto-encoding variational bayes,'' \emph{arXiv preprint arXiv:1312.6114}, 2013.

\end{thebibliography}

\end{document}